\documentclass{sig-alternate}
\newcommand{\ignore}[1]{}
\usepackage[pass]{geometry}
\usepackage{fancyhdr}
\usepackage[normalem]{ulem}
\usepackage[hyphens]{url}
\usepackage{hyperref}
\usepackage{color}
\usepackage{soul}

\usepackage{graphicx}
\usepackage[caption=false,font=footnotesize]{subfig}
\usepackage{amsmath}

\usepackage{multirow}
\usepackage{booktabs}

\usepackage{cite}
\usepackage{enumitem}
\usepackage{comment}
\usepackage{algorithm}
\usepackage{algpseudocode}
\algnewcommand\algorithmicforeach{\textbf{foreach}}
\algdef{S}[FOR]{ForEach}[1]{\algorithmicforeach\ #1\ \algorithmicdo}
\usepackage{tikz}
\newcommand*\circled[1]{\tikz[baseline=(char.base)]{
             \node[shape=circle,draw,inner sep=1pt] (char) {#1};}}

\title{Remapped Cache Layout: Thwarting Cache-Based Side-Channel Attacks with a Hardware Defense}
\begin{document}

\numberofauthors{3}
\author{
  \alignauthor Wei~Song\titlenote{This paper was submitted to HPCA'19 and rejected. The work was finished between December 2017 and May 2018.}\\
  \affaddr{SKLOIS, Institute of Information Engineering}\\
  \affaddr{Chinese Academy of Sciences}\\
  \affaddr{School of Cyber Security}\\
  \affaddr{University of Chinese Academy of Sciences}\\
  \affaddr{Beijing, China}
  \alignauthor Rui~Hou\\
  \affaddr{SKLOIS, Institute of Information Engineering}\\
  \affaddr{Chinese Academy of Sciences}\\
  \affaddr{School of Cyber Security}\\
  \affaddr{University of Chinese Academy of Sciences}\\
  \affaddr{Beijing, China}
  \alignauthor Peng~Liu\\
  \affaddr{The Pennsylvania State University}\\
  \affaddr{University Park, USA}
  \and
  \alignauthor Xiaoxin~Li\\
  \affaddr{SKLOIS, Institute of Information Engineering}\\
  \affaddr{Chinese Academy of Sciences}\\
  \affaddr{Beijing, China}
  \alignauthor Peinan~Li\\
  \affaddr{SKLOIS, Institute of Information Engineering}\\
  \affaddr{Chinese Academy of Sciences}\\
  \affaddr{Beijing, China}
  \alignauthor Lutan~Zhao\\
  \affaddr{SKLOIS, Institute of Information Engineering}\\
  \affaddr{Chinese Academy of Sciences}\\
  \affaddr{Beijing, China}
  \and
  \alignauthor Xiaofei Fu\\
  \affaddr{SKLOIS, Institute of Information Engineering}\\
  \affaddr{Chinese Academy of Sciences}\\
  \affaddr{Beijing, China}
  \alignauthor Yifei~Sun\\
  \affaddr{SKLOIS, Institute of Information Engineering}\\
  \affaddr{Chinese Academy of Sciences}\\
  \affaddr{Beijing, China}
  \alignauthor Dan~Meng\\
  \affaddr{SKLOIS, Institute of Information Engineering}\\
  \affaddr{Chinese Academy of Sciences}\\
  \affaddr{Beijing, China}
}

\maketitle
%\thispagestyle{firstpage}
%\pagestyle{plain}

%%%%%% -- PAPER CONTENT STARTS-- %%%%%%%%

\begin{abstract}
As cache-based side-channel attacks become serious security problems,
various defenses have been proposed and deployed in both software and hardware.
Consequently, cache-based side-channel attacks on
processes co-residing on the same core are becoming extremely difficult.
Most of recent attacks then shift their focus to the last-level cache (LLC).
Although cache partitioning is currently the most promising defense against the attacks abusing LLC,
it is ineffective in thwarting the side-channel attacks
that automatically create eviction sets or bypass the user address space layout randomization.
In fact, these attacks are largely undefended in current computer systems.

We propose Remapped Cache Layout (\textsf{RCL}) --
a pure hardware defense against a broad range of conflict-based side-channel attacks.
\textsf{RCL} obfuscates the mapping from address to cache sets;
therefore, an attacker cannot accurately infer the location of her data in caches
or using a cache set to infer her victim's data.
To our best knowledge,
it is the first defense to thwart the aforementioned largely undefended side-channel attacks .
\textsf{RCL} has been implemented in a superscalar processor
and detailed evaluation results show that
\textsf{RCL} incurs only small costs in area, frequency and execution time.
\end{abstract}

\section{Introduction}\label{sec_intro}

For decades computers have been designed to run applications fast and energy-efficiently.
As one of the key enabling techniques, caches are utilized to bring recently and frequently used data close to cores.
They are scarce microarchitectural resources which are dynamically shared between processes, cores and virtual machines (VMs).
To exploit the full benefits of caches,
they are mostly self-managed~\cite{Jacob2008} in a way that address space isolation is relaxed and traded for cache efficiency.
As a result, data from different processes, cores and VMs can reside in the same cache set and interfere with each other.
This interference opens a gate for various types of cache-based side-channel attacks.

As cache-based side-channel attacks become serious security problems for current computers,
various defenses have been proposed and deployed in both software and hardware.
Thanks to these defenses, cache-based side-channel attacks on processes co-residing on the same core
are becoming extremely difficult.
%Recent cryptography standards, such as AES,
%are implemented in constant-time programs~\cite{Bernstein2012}
%to remove the key dependent time variances.
It is observed that, modern operating systems (OSs) disable symmetric multithreading (SMT)
to mainly thwart side-channel attacks on L1 caches~\cite{Percival2005, Osvik2006, Aciicmez2007}.
Cloud providers also disallow sharing cores between VMs to prevent cross-VM attacks through L1 caches~\cite{Zhang2012}.
Additionally, flushing local caches (including L1) during process switching appears to
be a strong defense~\cite{Domnitser2012, Zhang2013a}.

Most of recent cache-based side-channel attacks shift their focus to the last-level cache (LLC)
because it is shared by processes running on different cores.
The \texttt{cflush} instruction of Intel x86-64 has been extensively exploited in flush-based cross-core attacks
to manipulate the memory shared between processes~\cite{Yarom2014, Gruss2016b}.
In the case where memory sharing is impossible,
attackers can launch conflict-based attacks by triggering accurate cache eviction inside LLC~\cite{Inci2016, Disselkoen2017}.
Furthermore, advanced attackers are capable of automatically creating eviction sets~\cite{Liu2015},
locating the security-critical data~\cite{Kayaalp2016},
bypassing the address space layout randomization (ASLR)~\cite{Gruss2016a, Gras2017},
and breaking the Intel SGX or ARM TrustZone isolation~\cite{Haehnel2017, vanBulck2017}.

Cache partitioning~\cite{Liu2016} is currently the most promising defense against attacks abusing LLC.
Cache partitioning separates security-critical data from normal data; therefore,
attackers cannot flush security-critical data or evict them using normal data.
However, cache partitioning is ineffective when security-critical data cannot be
easily separated from normal data~\cite{Gruss2016, vanBulck2017}
or normal data become the attacking targets~\cite{Gruss2016a, Gras2017}.
As a result, cache partitioning fails to thwart attacks
that automatically create eviction sets~\cite{Liu2015} or bypass the user ASLR~\cite{Gras2017},
leaving them larged undefended in current computer systems.

In this paper, we propose Remapped Cache Layout (\textsf{RCL}) --
a pure hardware defense against a broad range of conflict-based side-channel attacks.
Unlike cache partitioning which seeks to separate security-critical data from normal data,
\textsf{RCL} obfuscates the mapping from address to cache sets;
therefore, an attacker cannot accurately infer the location of her data in caches
or using a cache set to infer her victim's data.
To our best knowledge, \emph{\textsf{RCL} is the first defense to thwart the aforementioned largely undefended attacks.}
\textsf{RCL} is also a pure hardware defense which requires no software involvement.
In summary, the main contributions of this paper are as follows:
\begin{itemize}[noitemsep,nolistsep]
\item A pure hardware defense against a wide range of conflict-based side-channel attacks,
      namely Remapped Cache Layout (\textsf{RCL}), is proposed.
\item A systematic review of the common cache-based side-channel attacks
      and the existing defenses shows that
      three types of advanced attacks are largely undefended in current computer systems.
\item A detailed analysis reveals how \textsf{RCL} thwarts these three types of attacks.
\item \textsf{RCL} has been implemented in a superscalar processor
      and the evaluation results show that \textsf{RCL} incurs only small costs
      in area, frequency and execution time.
\end{itemize}

\begin{table*}[bt]
  \caption{Classification of common cache-based side-channel attacks}\label{tab_attack_class}
  \small{
  \begin{center}  
    \begin{tabular}{lcccll}
      \toprule
      Attacks         & Method       & Cache & Addr.  &  Vulnerability               & Note   \\
      \toprule
      \textbf{A-1}:~\cite{Percival2005, Neve2006, Osvik2006, Aciicmez2007, Zhang2012}
                      & Prime+Probe  & L1   & VA      & SMT                        & Co-resident processes. \\
      \textbf{A-2}:~\cite{Irazoqui2015, Liu2015, Inci2016, Kayaalp2016, Lipp2016, Disselkoen2017}
                      & Prime+Probe  & LLC  &  VA     & LgPg                       & Cross cores. \\
      \midrule
      \textbf{B}:~\cite{Gullasch2011, Irazoqui2014, Yarom2014, Zhang2014b, Gruss2015, Allan2016, Gruss2016b, Lipp2016, Zhang2016a}
                      & Flush+Reload & LLC  &  VA     & Dedup, SLib, PgMap         & Cross cores. \\
      \midrule
      \textbf{C-1}:~\cite{Liu2015, Oren2015, Gruss2016a}
                      & Prime+Reload & LLC  & VA      & LgPg, JS API               & Auto eviction set. \\
      \textbf{C-2}:~\cite{Maurice2015, Gruss2016a, Schwarz2017}
                      & Prime+Reload  & LLC   &  VA    & RowBuf, Pfc               & Recover PA. \\
      \midrule
      \textbf{D-1}:~\cite{Gras2017} 
                      & Evict+Time  & L1    & VA      & cached PTE                 & Bypass user ASLR. \\
      \textbf{D-2}:~\cite{Hund2013, Gruss2016} 
                      & Flush/Prefetch & L1 & VA      & MMU time-channel           & Bypass kernel ASLR. \\
      \midrule
      \textbf{E}:~\cite{Xu2015, Guanciale2016, Zhang2016, Haehnel2017, vanBulck2017}
                      & ---         & ---   &  PA     & malicious OS kernel        & SGX/TrustZone. \\
      \bottomrule
    \end{tabular}
  \end{center}
  \vspace{-0.8em}
  SLib: shared library, LgPg: larger page, Dedup: page deduplication, PgMap: \texttt{/proc/self/pagemap},
  RowBuf: DRAM rowbuffer, Pfc: performance counter, JS API: javascript application interface,
  MMU: memory management unit.
  \vspace{-0.5em}
  }
\end{table*}

\section{Caches in Modern Processors}\label{sec_cache}

In modern processors,
caches are utilized to store recently or frequently used data to reduce access time.
Data are stored in units of fixed-sized cache lines.
Caches are organized in a hierarchical structure.
Each core contains a pair of small but fast private L1 caches for data (L1-D) and instruction (L1-I).
The core may contain a medium-sized and unified level-two (L2) cache.
All cores share a large but comparatively slow LLC cache.
Commonly, caches are set-associative which allows a group of cache lines
to reside in one of the many cache sets.
Cache sets are addressed by an internal cache index,
which is typically a subset of the address bits shared by all cache lines in the same set.
When a new cache line is fetched but the corresponding cache set is full,
a cache replacement policy chooses and evicts a cache line from this set.

A typical L1 cache is addressed by the virtual address (VA),
while other levels of caches are addressed by the physical address (PA).
The VA to PA translation in L1 caches proceeds in parallel with cache set accesses.
A translation lookaside buffer (TLB) is used to reduce this translation latency.
The mapping from VA to PA is managed by the OS kernel in units of pages.
The whole mapping is recorded in a page table stored in memory.
The aforementioned TLB caches the recently accessed mapping
so that no actual access to the page table is needed for a recently accessed page.

An inclusive cache hierarchy is normally assumed,
where a cache line evicted from LLC is also purged from all cache levels.
A cache coherence protocol is utilized
to ensure data are correctly updated in all levels of caches.

\section{Attack Classification}~\label{sec_attack}

Cache-based side-channel attacks involve an attacker and a victim sharing a cache.
The attacker utilizes this cache to infer information belonging to the victim.
Table~\ref{tab_attack_class} provides a summary of common cache-based side-channel attacks.

\subsubsection*{\textbf{A}: Conflict-based side-channel attacks}\label{subsec_attack_a}

Conflict-based side-channel attacks~\cite{Yan2017} exploit the fact that
caches indiscriminately store congruent cache lines
(lines sharing the same subset of address bits)
in the same cache set.
This allows attackers to maliciously
prime certain cache sets using her own data (such as Prime+Reload~\cite{Percival2005})
or evict the victim from certain cache sets (such as Evict+Time~\cite{Osvik2006}).
According to the level of caches being targeted,
conflict-based side-channel attacks have two varieties:

\textbf{A-1}:
\emph{Conflict-based attacks targeting the co-resident processes sharing the L1 cache}.
Since the L1 cache is small and fast,
the attacker can extract detailed information with low noise and at high speed.
To synchronize with her victim, the attacker can utilize SMT to run concurrently with the victim~\cite{Percival2005, Osvik2006}
or preempt her victim at short intervals~\cite{Neve2006,Aciicmez2007}.
However, running attacking process on the same core with the victim can be difficult in multicore processors.
Most cloud service providers have disabled SMT and disallowed multiple VMs running on the same core.

\textbf{A-2}:
\emph{Conflict-based cross-core attacks targeting LLC.}
To remove the co-residence requirement,
a conflict-based attacks can be launched from a process running on another core or
even another VM~\cite{Irazoqui2015, Liu2015, Inci2016, Lipp2016, Disselkoen2017}
through side-channels in LLC.
What is worse, the advanced attacks can automatically create eviction sets~\cite{Liu2015, Oren2015}
or discover security-critical data~\cite{Disselkoen2017}.
This is the by far the most sophisticated type of cache-based side-channel attacks
and is thus classified separately as type \textbf{C}.

\subsubsection*{\textbf{B}: Flush-based side-channel attacks}

Although conflict-based side-channel attacks are powerful and extremely adaptive,
when the target memory is shared between the attacker and her victim,
with the permission of the OS,
the attacker can easily launch side-channel attacks by flushing the cache lines of interest.
Unfortunately, the \texttt{cflush} instruction of the Intel x86-64 ISA deliberately allows
user applications to flush cache lines from LLC,
which becomes a notorious vulnerability utilized
by a family of attacks~\cite{Gullasch2011, Irazoqui2014, Yarom2014, Zhang2014b, Gruss2015, Allan2016, Gruss2016b, Lipp2016, Zhang2016a}.
As a prerequisite, the target memory must be shared between the attacker and her victim.
Shared libraries have been used to attack processes running in the same OS~\cite{Zhang2014b, Gruss2015, Lipp2016}.
The physical page mapping exposed by some OSs (\texttt{/proc/self/pagemap})~\cite{Gruss2016, Lipp2016}
and memory deduplication have been used to launch cross-VM attacks~\cite{Irazoqui2014, Yarom2014, Gruss2015, Allan2016, Gruss2016b},
although both vulnerabilities have been neutralized in latest OSs and VMs.

\subsubsection*{\textbf{C}: Automatic side-channel attacks}\label{subsec_attack_c}

As mentioned in \textbf{A-2}, advanced conflict-based attacks become adaptive at run-time.

\textbf{C-1}:
\emph{Automatically creation for eviction sets.}
When an attacker runs in non-pointer environment, she cannot even obtain the VAs of her own variables.
To launch conflict-based attacks targeting LLC, attacker must create eviction sets to trigger accurate cache evictions.
For this purpose,
a large amount of memory is acquired using system APIs~\cite{Oren2015, Gruss2016a} or larger pages~\cite{Liu2015}.
A blinded but optimized search algorithm~\cite{Liu2015} is then used to create eviction sets at run-time.

\textbf{C-2}:
\emph{Reversing the VA to PA mapping.}
Furthermore, attackers can mobilize the eviction sets created by \textbf{C-1} to exploit other side-channels,
such as the row buffer in current DRAM chips.
By exploiting the PA invariance related to the row buffer,
the attacker can infer a part of the PA of the created eviction set~\cite{Maurice2015, Gruss2016a, Schwarz2017}.

\begin{table*}[bt]
  \caption{Existing hardware defenses against cache-based side-channel attacks}\label{tab_defense_class}
  \small{
  \begin{center}  
    \begin{tabular}{lccccccccc}
      \toprule
                          &             & \multicolumn{8}{c}{Protection level against attack sub-types}               \\
      \cmidrule(r){3-10}
      Defenses            & Software modification & \textbf{A-1} & \textbf{A-2}
                                                       & \textbf{B}   & \textbf{C-1}
                                                                      & \textbf{C-2} & \textbf{D-1}
                                                                                     & \textbf{D-2} & \textbf{E} \\
      \toprule
%                                                      A-1 A-2  B    C-1         C-2          D-1          D-2  E
      Cache partitioning~\cite{Page2005, Wang2007, Wang2008, Domnitser2012, Liu2016, Yan2017}
                          & Yes (set/way allocation)  & S & S & S & \textbf{W} & \textbf{W} & \textbf{W} & W & M \\
      \midrule
      Random (pre-)fetch/decay~\cite{Keramidas2008, Liu2014, Fuchs2015}
                          & Might (set fetch region)  & M & M & S & \textbf{W} & \textbf{W} & \textbf{M} & M & M \\
      \midrule
      Random cache layout~\cite{Wang2007, Wang2008}
                          & Might (set way priority)  & S & W & W & \textbf{W} & \textbf{W} & \textbf{M} & W & W \\
      \midrule
      Software isolation~\cite{McKeen2013, Costan2016}
                          & Yes (enclave management)  & S & S & S & \textbf{W} & \textbf{W} & \textbf{W} & W & S \\
      \bottomrule
    \end{tabular}
  \end{center}
  \vspace{-0.8em}
  SW: software, S: strong, M: medium, W: weak.
  \vspace{-0.5em}
  }
\end{table*}

\subsubsection*{\textbf{D}: Page table side-channel attacks}

Sine modern processors cache the recently accessed page table entries (PTEs) in L1-D cache,
caches are exploited to break the ASLR protection.

\textbf{D-1}:
\emph{Bypassing the user ASLR.}
If running in a sandboxed environment, the virtual memory space is normally randomized by ASLR.
As the first step to jailbreak the sandbox,
an attacker can launch conflict-based attacks on the cached PTEs to bypass ASLR~\cite{Gras2017}.

\textbf{D-2}:
\emph{Bypass the kernel ASLR.}
ASLR has also been used to randomize the kernel memory space
as a first defense against code-reuse attacks~\cite{Snow2013, Gruss2017}.
To break the kernel ASLR, attackers can flush or prefetch the targeted PTE from the L1-D cache and then reload the target page.
A timing difference could be detected if the page exists~\cite{Hund2013, Gruss2016}.

\subsubsection*{\textbf{E}: Privileged side-channel attacks}

If an OS kernel is malicious, the kernel can launch direct side-channel attacks on trusted execution environment (TEE)~\cite{McKeen2013},
such as the Intel SGX~\cite{Xu2015, Haehnel2017, vanBulck2017} and the ARM Trustzone~\cite{Guanciale2016, Zhang2016}.
These attacks normally exploit the facts
that TEEs still rely on the kernel to allocate memory
and a malicious OS kernel can launch flush-based cache attacks using PAs rather than VAs,
which significantly increases accuracy.

\section{Existing defenses}\label{sec_defense}

Various defenses have been proposed using software, hardware or together.
For the various types of attacks,
some types are better resolved in software
but many are better handled by hardware
to avoid software modification and severe performance loss.

On the software side, most cryptography standards are hardened with constant-time programming~\cite{Bernstein2012}
utilizing the cryptographic instruction extensions supported by major processors~\cite{Gueron2010,Gouvea2015}.
Furthermore, enforcing the flushing of local caches (including L1 caches)~\cite{Domnitser2012, Zhang2013a} during process switching
appear to be effective for all cross-process attacks targeting L1 caches.
Deploying the Intel hardware transnational memory support to enforcing the preloading of security-critical data upon context switch~\cite{Gruss2017a}
also shows strong defense against all cross-process attacks.
Optimistically speaking, cache-based side-channel attacks targeting co-resident processes sharing the L1 cache (sub-type \textbf{A-1})
can be effectively handled by software defenses.

For conflict-based cross-core attacks targeting LLC (sub-type \textbf{A-2}),
using page color~\cite{Shi2011} or explicit management on LLC~\cite{Kim2012, Liu2016}
to achieve cache partitioning has been proven effective.
These defenses normally require changes in hypervisors.
Explicit cache management also compromises the self-management of caches,
which may lead to significant performance loss. 

The flush-based cross-core attacks (type \textbf{B}) rely on the availability of flush and prefetch instructions,
and the sharing of the target memory between the attacker and her victim.
A majority of these attacks can be effectively resolved by constraining the flush and prefetch instructions~\cite{Yan2017},
denying unprivileged accesses to page maps (\texttt{/proc/self/pagemap})~\cite{Gruss2016},
and removing the memory deduplication in hypervisors~\cite{Gruss2016b}.

The attacks trying to bypass the kernel ASLR (sub-type \textbf{D-2}) can be effectively
resolved by separating the kernel virtual space from the user virtual space~\cite{Gruss2017},
namely the kernel page table isolation (KPTI).

On the hardware side, Table~\ref{tab_defense_class} summarizes the existing hardware defenses.
\emph{Cache partitioning} can be enforced in hardware as well.
Both software and hardware cache partitioning can effectively thwart cross-process attacks.
The hardware enforced partitioning reduces the burden of software~\cite{Domnitser2012, Yan2017} but may still require software
to manage the cache set/way allocation~\cite{Page2005, Wang2007, Wang2008, Liu2016}.
Cache partitioning is powerless when the victim's cache lines cannot be separated from the attacker's (type \textbf{C} and \textbf{D}).
It is also ineffective when the OS kernel is malicious (type \textbf{E})
because the attacker may directly manipulate the set/way allocation~\cite{Page2005, Wang2007, Wang2008, Liu2016}.

\emph{Randomly (pre-)fetching or decaying cache lines}~\cite{Keramidas2008, Liu2014, Fuchs2015}
introduces noise to the flush/prefetch operation.
It thus provides a relatively strong defense against flush-based attacks (type \textbf{B}). 
Similarly, it introduces noise to the conflict-based attacks.
However, sophisticated conflict-based attacks (type \textbf{C}) always average out the noise by repeating the attack procedure.
Meanwhile, defenses may require software to setup the random fetching~\cite{Liu2014}.

\emph{Randomizing the cache layout}~\cite{Wang2007, Wang2008}
has been proven a strong defense against conflict-based attacks.
It prevents attackers from easily inferring cache sets from addresses and vice versa
by deliberately randomizing the mapping from addresses to cache sets.
It has recently been utilized in a browser (software) to protect the user ASLR~\cite{Schwarz2018}.
However, the randomization provided by existing hardware defenses~\cite{Wang2007, Wang2008} is limited.
It is not applicable to LLC and congruent cache lines are still stored in the same cache set.
These make the existing defenses incapable of handling conflicted-based attacks targeting LLC (sub-type \textbf{A-2})
or attacks utilizing the automatically created eviction sets (type \textbf{C}).

\emph{Software isolation} provides a TEE
for running security-critical application~\cite{McKeen2013, Costan2016}.
This is the recommended method to protect applications from malicious OSs (type \textbf{E}).
However, it fails to stop an attacker from actually utilizing an enclave to
launch and conceal conflict-based attacks (type \textbf{C})~\cite{Schwarz2017}.

In summary,
software defenses can effectively thwart attacks of types \textbf{A}, \textbf{B} and \textbf{D-2}.
Hardware defenses can effectively thwart attacks of types \textbf{A}, \textbf{B} and \textbf{E}.
These leave the attacks of types \textbf{C} and \textbf{D-1} largely undefended in current computer systems.
To thwart these attacks, this paper proposes \textsf{RCL},
which also randomizes the cache layout
but does not suffer from the limited randomization
as the previous defenses~\cite{Wang2007, Wang2008} does.
A comparison will be provided in Section~\ref{sec_related}.

\section{Remapped Cache Layout}

As described in Section~\ref{subsec_attack_a}.A,
conflict-based side-channel attacks exploit the fact that caches indiscriminately store
congruent cache lines in the same cache set.
To thwart conflict-based attacks, we propose Remapped Cache Layout (\textsf{RCL}),
which deliberately randomizes the mapping from addresses to cache sets.
In the view of attackers, irrelevant cache lines are stored in the same cache set;
therefore, they cannot deterministically create eviction sets.
Even if eviction sets are created using an automatic searching algorithm~\cite{Liu2015},
it is extremely difficult to figure out which cache sets are evicted.

In a normal $w$-way $2^s$-set L1-D cache,
the cache index $CI$ comes from the lower $s$ bits of the VA\footnote{Here we assume a cache line is 64 bytes and a page is 4~KiB.}:
\begin{equation}\label{eqn_ci_normal}
CI = VA[s+5:6]
\end{equation}
This direct mapping from VAs to cache sets (PA for other cache levels) is
the fundamental microarchitectural vulnerability enabling all conflict-based attacks. 

\begin{figure}[bt]
\centering{
\includegraphics[width=0.43\textwidth]{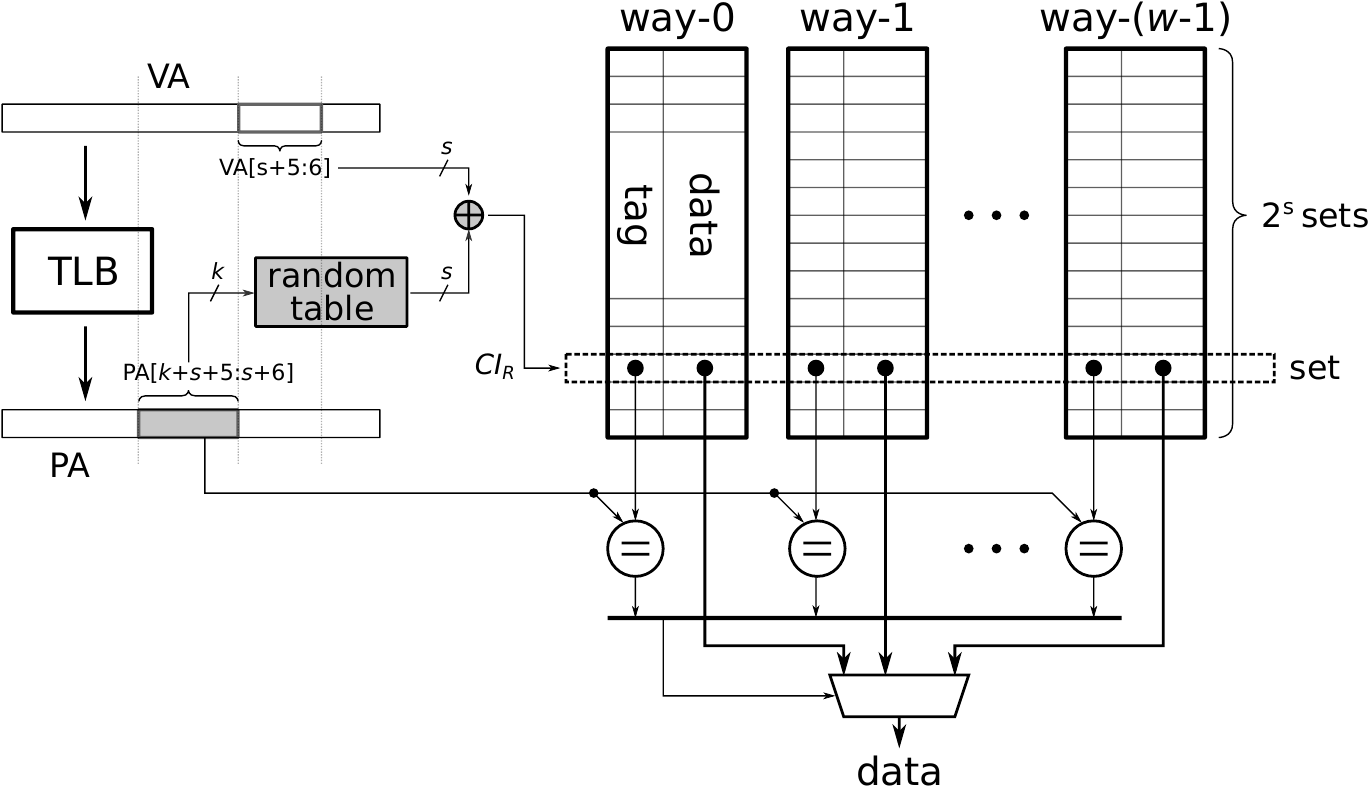}
\vspace{-0.8em}
}
\caption{Applying \textsf{RCL} on a L1 cache}\label{fig_rcl_l1d}
\vspace{-0.5em}
\end{figure}

In a \textsf{RCL} enabled L1 cache, as depicted in Figure~\ref{fig_rcl_l1d},
the randomized cache index $CI_R$ is calculated from both VA and PA:
\begin{equation}\label{eqn_ci_rcl_l1d}
  CI_R = RT(PA[k+s+5:s+6]) \oplus VA[s+5:6]
\end{equation}
Here $k$ is a implementation defined parameter tuning the level of introduced randomness.
\textsf{RCL} picks $k$ bits from PA ($PA[k+s+5:s+6]$)
and uses it as an index to obtain an $s$-bit true random number from the random table (RT).
This random number is then XORed with the original cache index $VA[s+5:6]$.

As shown in Table~\ref{tab_attack_class}, except for type \textbf{E},
all types of attacks do not have the direct access to PA.
Assuming all the software vulnerabilities (such as \texttt{/proc/self/pagemap}~\cite{Gruss2016})
that expose PA to user applications are removed,
which is true for most modern computer systems,
PA is a secret hidden behind MMU.
However, PA is not a perfect source for randomness.
Assuming the attacker is able to acquire a large amount of physically and virtually consecutive memory,
such as a larger page~\cite{Liu2015},
the linear pattern existing in PA is vulnerable to brute-force attacks.

The random table depicted in Figure~\ref{fig_rcl_l1d} is added to enhance the randomness introduced from PA.
This random table contains $2^k$ true random numbers, each of which is $s$ bits wide.
It guarantees that two ($CI_R$)s  are mutually independent even
there is only a single bit difference in $PA[k+s+5:s+6]$.
To introduce the maximum randomness derived from PA,
the lowest $k$ bits in the part of PA non-overlapped with the VA are used to index the random table.
This parameter $k$ also defines the coverage of the layout randomization.
The cache lines in the address range of $2^{k+s+6}$ are randomly mapped to all available cache sets.

For other levels of caches, such as LLC,
the original cache index is $PA[s+5:6]$ as the cache is physically addressed.
Accordingly, the cache index for a \textsf{RCL} enabled LLC is calculated from PA:
\begin{equation}\label{eqn_ci_rcl_llc}
  CI_R = RT(PA[k+s+5:s+6]) \oplus PA[s+5:6]
\end{equation}

Choosing PA as the source of randomness is deliberate.
\textsf{RCL} is a pure and modular hardware defense which should be independently deployed to all cache levels for the beast protection.
Neither software observable (at least for user applications) nor LLC invisible information should be used.
PA is the only hardware information
that is both widely available to all levels of caches and hidden from user applications.

\textsf{RCL} is a pure hardware defense because both PA and the random table
are purely hardware managed information.
As a result, \textsf{RCL} requires no modification in any levels of software.
It can be deployed in commercially ready processors without disrupting the existing software ecosystem.

\section{Security Analysis of \textsf{\large RCL}}\label{sec_ana}

Table~\ref{tab_rcl_protection} summarizes the effectiveness of \textsf{RCL} related to
the common cache-based side-channel attacks listed in Table~\ref{tab_attack_class}.
\textsf{RCL} thwarts most attacks that need to evict cache sets.
Most importantly, to our best knowledge,
\emph{\textsf{RCL} is the first defense to thwart the automatic side-channel attacks (type \textbf{C})
and the attacks bypassing the user ALSR (sub-type \textbf{D-1}).} 

\begin{table}[bt]
  \caption{Effectiveness of \textsf{RCL}}\label{tab_rcl_protection}
  \small{
  \begin{center}  
    \begin{tabular}{lcl}
      \toprule
      Attacks           & Protection         & Existing defenses\\
      \toprule
      \textbf{A-1}      & Strong             & Flush L1 caches~\cite{Domnitser2012, Zhang2013a}\\
      \textbf{A-2}      & Strong             & Cache partitioning~\cite{Domnitser2012, Yan2017}\\
      \midrule
      \textbf{B}        & Weak               & Constrain flush/prefetch~\cite{Yan2017}\\
      \midrule
      \textbf{C-1}      & \textbf{Strong}    & \textbf{None} \\
      \textbf{C-2}      & \textbf{Strong}    & \textbf{None} \\
      \midrule
      \textbf{D-1}      & \textbf{Strong}    & \textbf{None} \\
      \textbf{D-2}      & Weak               & KPTI~\cite{Gruss2017} \\
      \midrule
      \textbf{E}        & Weak               & Software isolation (TEE)~\cite{Costan2016}\\
      \bottomrule
    \end{tabular}
  \end{center}
  }
  \vspace{-0.5em}
\end{table}

\subsubsection*{\textbf{A}: Conflict-based side-channel attacks}\label{subsec_ana_a}

\textsf{RCL} is effective in thwarting all conflict-based side-channel attacks
when the VA to PA translation is not exposed and
the VAs of security-critical data are not leaked by other means.
The strong protection of \textsf{RCL} comes from three reasons:
\begin{itemize}[noitemsep,nolistsep]
\item \emph{\textsf{RCL} significantly increases the difficulties in creating eviction sets.}
In normal conflict-based attacks,
attackers utilize the direct mapping revealed in Equation~\ref{eqn_ci_normal}
to deterministically construct eviction sets containing congruent cache lines.
Without knowing the PA of her own data and the content of the random table,
attackers are incapable of constructing eviction sets in the usual way.
The only available to create eviction sets is the automatic search algorithm~\cite{Liu2015}
as described in Section~\ref{subsec_attack_c}.C.
\item \emph{\textsf{RCL} significantly reduces the amount of information leaked by cache eviction.}
Even eviction sets are created using the automatic search algorithm,
as it will be revealed in Section~\ref{subsec_ana_c}.C,
\textsf{RCL} introduces significant amount of noises when they are applied to LLC.
\item \emph{Ultimately, \textsf{RCL} removes the spatial correlation between an attacker and her victim through a cache.}
Even if eviction sets are created by a yet unknown means and there is a yet unknown way to reduce the noise introduced by \textsf{RCL},
an attacker still needs to infer useful information using a certain cache set as a medium.
Since the cache layout is randomized, without disclosing the PA and deciphering the random table,
there is no deterministic way to spatially correlate the data of any two pages.
\end{itemize}

\begin{figure}[bt]
\centering{
\subfloat[\textsf{RCL} cache ($k=6$))]{\includegraphics[width=0.43\textwidth]{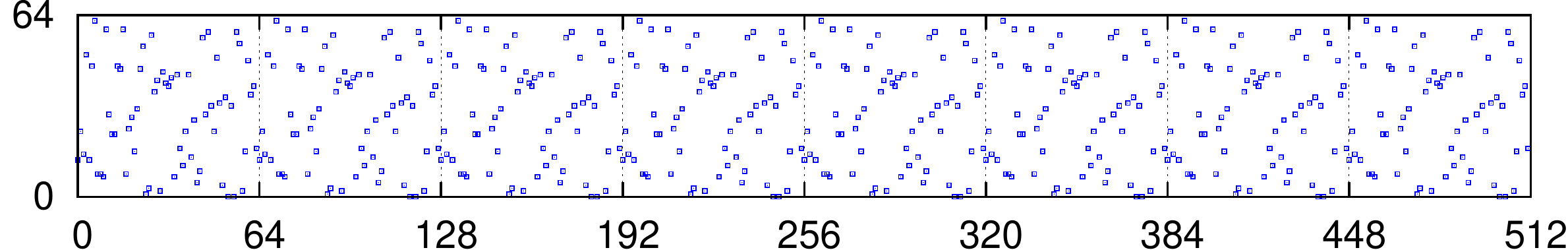}\label{fig_bitmap_randtab}}\\
\vspace{-0.5em}
\subfloat[\textsf{RCL} cache ($k=9$)]{\includegraphics[width=0.43\textwidth]{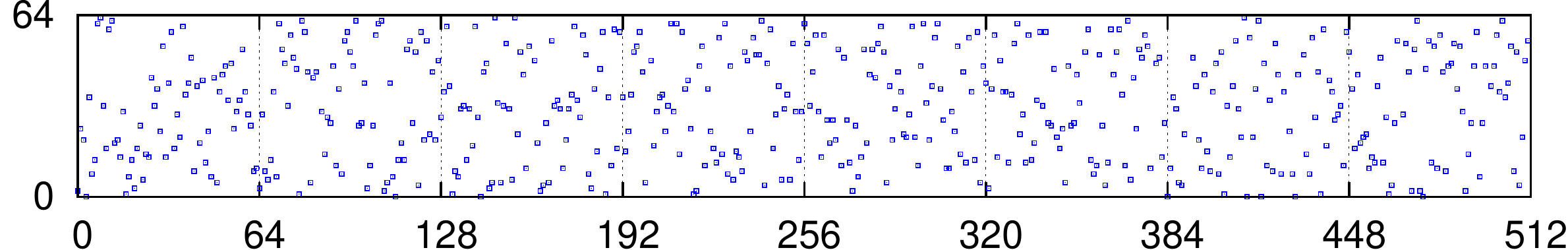}\label{fig_bitmap_randtab_ext}}
}
\vspace{-0.5em}
\caption{Cache lines mapped to the same set}\label{fig_bitmap}
\vspace{-0.5em}
\end{figure}

For the conflict-based attacks targeting L1 caches (sub-type \textbf{A-1}),
if an attacker can successfully acquire a physically and virtually consecutive memory space, such as a larger page,
larger than $w \cdot 2^{k+s+6}$ bytes,
the attacker is able to deterministically create eviction sets
as any cache lines sharing the same $PA[k+s+5:6]$ are still mapped to the same cache set.
Figure~\ref{fig_bitmap_randtab} shows a bitmap representing the cache lines inside a larger page (2~MiB) mapping to the same
set in a \textsf{RCL} enabled cache ($s=6, k=6$),
where the x axis denotes pages and the y axis denotes the cache lines in each page.
The distribution of cache lines is randomized only in the 64 pages space and then duplicated 8 times in the larger page,
which is enough to form an eviction set for a 4-way cache.
However, without knowing the exact PA and the content of the random table,
the attacker cannot know which cache set is evicted.
No existing attacks (targeting L1 caches) works in this way.
Nevertheless, if we are determined to eliminate this threat,
an implementation can increase $k$ to make $w \cdot 2^{k+s+6}$ bytes larger than a larger page (at the cost of a larger random table).
As shown in Figure~\ref{fig_bitmap_randtab_ext}, 
the distribution of cache lines is totally randomized when $k$ is increased to 9.
Alternatively, the system can disable the larger page support for user applications~\cite{Irazoqui2015}.
Flushing the L1 caches during process switching~\cite{Domnitser2012, Zhang2013a} is another way to effectively thwart any cache-base side-channel attacks targeting L1 caches.

For the conflict-based attack targeting LLC (sub-type \textbf{A-2}),
since the number of cache sets are significantly larger than that in L1 caches
and the cache is addressed by PAs rather than VAs,
creating eviction sets becomes even more difficult than doing it in L1 caches.
\textsf{RCL} breaks the direct mapping from addresses to cache sets.
All existing means to deterministically create cache sets cease to work.
For advanced attacks that create eviction sets using an automatic search algorithm (sub-type \textbf{C-1}),
\textsf{RCL} also provides a strong protection.
Detailed analysis will be provided later in Section~\ref{subsec_ana_c}.C related to sub-type \textbf{C-1}.

\subsubsection*{\textbf{B}: Flush-based side-channel attacks}

The flush and prefetch operations explicitly evict or load a cache line in the cache system using VAs.
These are legal cache management operations which will succeed no matter where the cache line is stored.
As a result, randomizing the cache layout has no effect.
\textsf{RCL} cannot stop flush-based side-channel attacks.
Fortunately, flush-based attacks can be effectively resolved
by constraining the flush and prefetch instructions
and removed the potential vulnerable memory sharing mechanisms
as described in Section~\ref{sec_defense}.

\subsubsection*{\textbf{C}: Automatic side-channel attacks}\label{subsec_ana_c}

For conflict-based attacks targeting LLC,
especially those attacking the post Sandy Bridge Intel processors
which have LLC sliced and addressed using the \emph{complex addressing} scheme~\cite{Maurice2015},
attackers rely on a blind but optimized search algorithm to create eviction sets for LLC (sub-type \textbf{C-1}).
Most automatic search algorithms are derived from Liu's algorithm~\cite{Liu2015}.
The input is a large set of cache lines
containing enough lines to evict a cache set.
By iteratively trimming irrelevant cache lines,
the search algorithm produces a comparatively pure eviction set targeting LLC.
This search algorithm is extremely robust.
It is ignorant to the complex addressing scheme,
appears to work in inclusive or non-inclusive cache hierarchies,
and even shows resistance to \textsf{RCL}.

\begin{figure}[bt]
\centering{
\subfloat[Normal cache]{\includegraphics[width=0.18\textwidth]{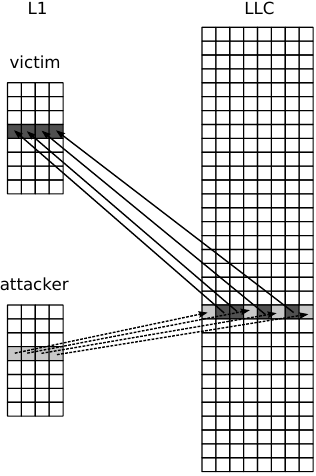}\label{fig_eviction_set_normal}} \qquad
\subfloat[\textsf{RCL} enabled LLC]{\includegraphics[width=0.18\textwidth]{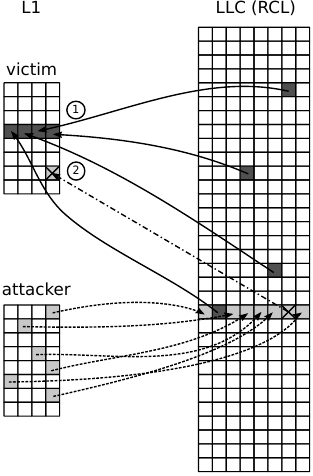}\label{fig_eviction_set_rcl}}
\vspace{-0.8em}
}
\caption{Automatic search for eviction sets}\label{fig_eviction_set}
\vspace{-0.5em}
\end{figure}

Figure~\ref{fig_eviction_set_normal} shows an eviction set created in a two-level inclusive cache hierarchy.
Utilizing this eviction set, the attacker is able to prime a LLC cache set.
Thanks to the direct mapping from addresses to cache sets and the inclusive cache hierarchy,
evicting a set in LLC effectively
evicts a large portion of a cache set \footnote{In Figure~\ref{fig_eviction_set_normal},
the whole set in the victim's L1 cache is evicted for simplicity.} inside the victim's L1 cache.

Now let us apply \textsf{RCL} to LLC.
\textsf{RCL} may fail to prevent Liu's algorithm from producing an eviction set.
Assuming this eviction set is pure, as shown in Figure~\ref{fig_eviction_set_rcl},
loading lines in this eviction set does evict a cache set in LLC
but they are distributed across arbitrary cache sets in the attacker's L1 cache.

Although \textsf{RCL} cannot prevent the creation of eviction sets,
it significantly reduces the usefulness of these eviction sets.
This reduction comes from two reasons:
\circled{1} The lines of a target cache set in the victim's L1 cache are scattered across a large number of cache sets in LLC.
Evicting a set in LLC most likely purges only one cache line from the target cache set in the victim's L1 cache.
\circled{2} It is likely that a set in LLC contains lines belonging to multiple cache sets in the victim's L1 cache.
Evicting the victim's cache lines through priming LLC is significantly noisy.
For these two reasons, most conflict-based side-channel attacks relying on the automatic creation of eviction sets stop to work.

Some advanced attacks might combine cache side-channels with other side-channels to reverse part of the PA (sub-type \textbf{C-2}).
However, these attacks either still rely on the direct mapping from address to cache sets in LLC~\cite{Maurice2015, Gruss2016a},
or requiring precisely evict two cache lines from LLC~\cite{Schwarz2017}.
The former no longer works if \textsf{RCL} is applied to LLC.
For the latter, a randomized cache replacement policy would significantly reduce the required precision.
Alternatively, oblivious RAM~\cite{Stefanov2013} would provide a strong protection against such attacks.
Other sophisticated attacks might use cache set collision to locate the security-critical information stored in the victim's L1 cache.
However, the actual information leak relies on the correlation between cache indexes and page offset~\cite{Kayaalp2016}.
Such correlation is removed by \textsf{RCL}.
In summary, \emph{\textsf{RCL} is effective in thwarting the automatic side-channel attacks,
which is so far the most sophisticated cache-based side-channel attacks.}

\subsubsection*{\textbf{D}: Page table side-channel attacks}

\textsf{RCL} easily thwart any attacks trying to bypass the user ASLR protection (sub-type \textbf{D-1}).
These attacks run in restricted environment (normally sandboxed) where an attacker cannot obtain the VA of her own variable~\cite{Gras2017}.
The attacker launches accurate cache evictions to locate the cached PTEs related to a VA
and then infer part of the VA using the cache indexes of these PTEs.
When \textsf{RCL} is applied, the correlation between the cache indexes of PTEs and the VA is safely removed.

For attacks trying to bypass the kernel ASLR protection (sub-type \textbf{D-2}),
\textsf{RCL} can thwart the attacks accurately evicting PTEs from the L1-D cache.
However, \textsf{RCL} cannot stop the attacks that flush/prefetch the PTEs in the L1-D cache~\cite{Gruss2016}
or the attacks exploiting the caches inside the memory management unit (MMU)~\cite{Hund2013} (translation and page table caches~\cite{vanSchaik2017}).
Nevertheless, the thorough solution is adopt KPTI to separate the kernel page table from user's~\cite{Gruss2017}.

\subsubsection*{\textbf{E}: Privileged side-channel attacks}

For the attacks disguising themselves in OSs,
the translation from VA to PA is already exposed
and attackers may have explicit control over the cache content (flush or prefetch).
Obscuring the cache layout, such as \textsf{RCL}, provides little protection.
A TEE armed with strong resource isolation~\cite{Costan2016} provides better protection.

\section{Implementation}\label{sec_imp}

This section demonstrates the implementations of \textsf{RCL}
in the open-source BOOM system-on-chip (SoC)~\cite{Celio2015, Asanovic2016}.
The BOOM processor is an open-source 6-stage out-of-order execution superscalar processor
running the 64-bit RISC-V ISA~\cite{Waterman2014}.
Every BOOM processor has a pair of private L1-D and L1-I caches
while all processors share a unified L2 cache (LLC) with coherence support.
We choose this open-source BOOM SoC because
it allows us to implement and evaluate \textsf{RCL} an actual and Linux-ready SoC.

\subsection{\textsf{RCL} in L1 caches}\label{subsec_imp_rcl_l1}

\begin{figure}[bt]
\centering{
\includegraphics[width=0.28\textwidth]{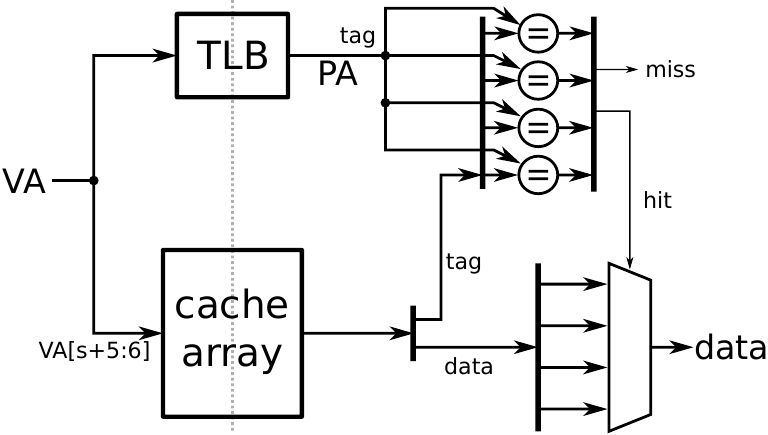}
\vspace{-0.8em}
}
\caption{A normal L1 cache}\label{fig_rcl_l1_normal}
\vspace{-0.5em}
\end{figure}

Figure~\ref{fig_rcl_l1_normal} depicts the implementation of a L1 cache in the BOOM processor.
It is pipelined and serves a memory read in two cycles if the requested data (instruction) is cached.
To reduce the cache access latency,
inquiry to the TLB and accessing the cache array proceed in parallel within the same clock cycle.

\begin{figure}[bt]
\centering{
\includegraphics[width=0.33\textwidth]{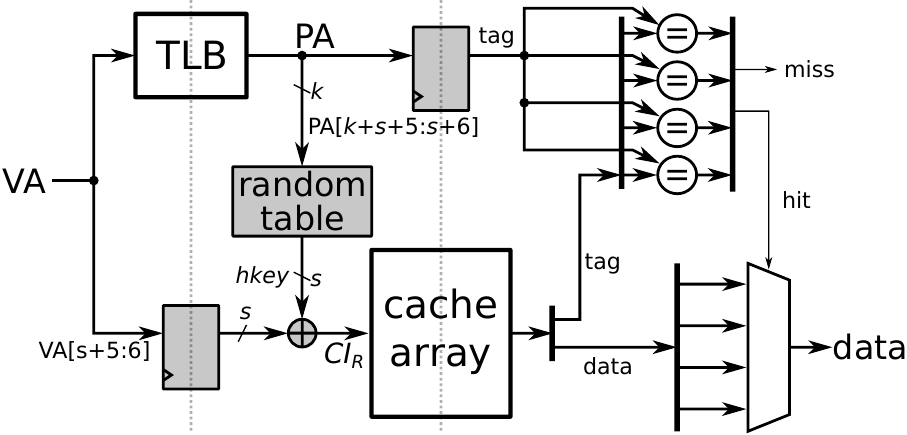}
\vspace{-0.8em}
}
\caption{Implementing \textsf{RCL} in an L1 cache}\label{fig_rcl_l1_cycle}
\vspace{-0.5em}
\end{figure}

To implement \textsf{RCL} in such a cache,
the cache array is indexed by $CI_R$ as described in Equation~\ref{eqn_ci_rcl_l1d},
which is the original $s$-bit $CI$ XORed with an $s$-bit hash key ($hkey$) calculated from a part of the PA:
\begin{equation}\label{eqn_ci_hkey_l1d}
  hkey = RT(PA[k+s+5:s+6])
\end{equation}
This requires the PA to be translated before accessing the cache array.
As shown in Figure~\ref{fig_rcl_l1_cycle},
the TLB inquiry is thus made to operate one cycle before accessing the cache array.
This serialization increases the cache access latency by one cycle.
Section~\ref{subsec_fpga} will analyze this impact in details.

The random table contains $2^k$ true random numbers, each of which is $s$ bits wide.
The $hkey$ is an entry read from the random table using a $k$-bit index extracted from the PA ($PA[k+s+5:s+6]$).
The implementation of the random table is cache depended.
In L1 caches, it is normally implemented in registers to reduce the table access latency.
The size of the random table in a small L1 cache is limited;
therefore, implementing the table using registers (which is more area demanding than SRAMs) does not incur significant area overhead.
For other levels of caches, random tables are implemented in SRAMs for area efficiency.

\subsection{A speculative L1-D cache}

\begin{figure}[bt]
\centering{
\includegraphics[width=0.37\textwidth]{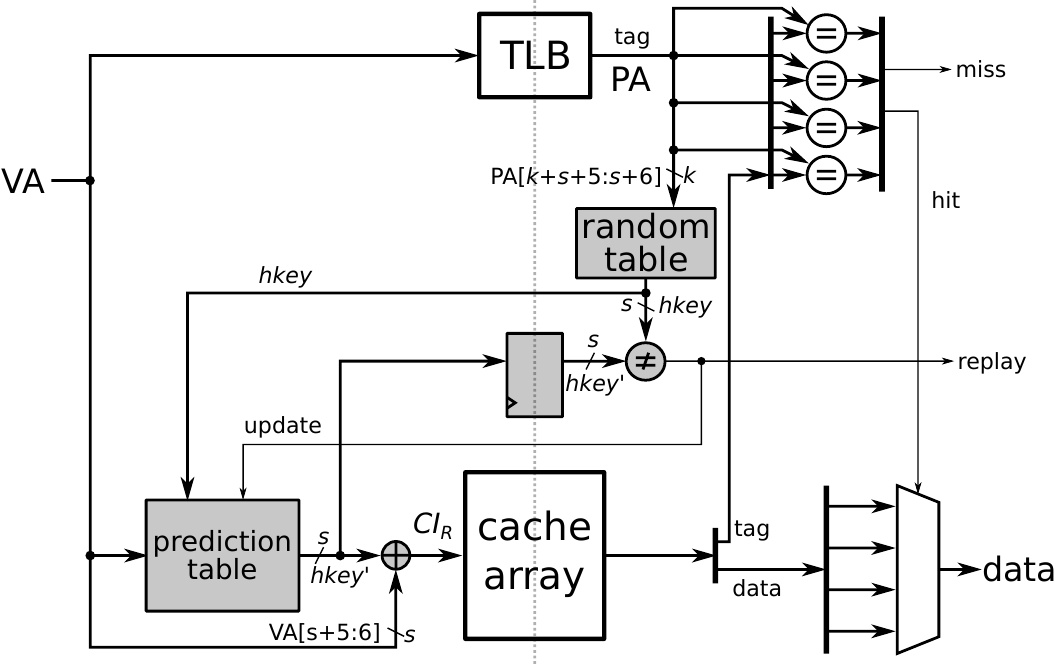}
\vspace{-0.8em}
}
\caption{A speculative \textsf{RCL} L1-D cache}\label{fig_rcl_l1d_speculate}
\vspace{-0.5em}
\end{figure}

As mentioned in Section~\ref{subsec_imp_rcl_l1},
the serialization of TLB and array access introduces an extra cycle for every cache access,
which can incur a noticeable performance penalty.
To avoid this extra cycle,
a cache can speculate a hash key ($hkey'$) using VA, as depicted in Figure~\ref{fig_rcl_l1d_speculate}.

The key component is the prediction table.
It is a small content addressable register file storing the hash keys of previously visited virtual pages.
Assuming the data access pattern present strong spatial locality,
most cache access are within a small number of pages and would hit in the prediction table.
Also since the table is small, it would not make the path critical in timing.

If a VA is missed in the prediction table,
such as when a new page is visited or a new page table is used due to context switch,
the speculated hash key would be wrong.
To detect such misprediction,
$hkey'$ is registered and compared with the correct hash key ($hkey$) in the next cycle.
If $hkey' \neq hkey$, denoting the wrong cache set is accessed in the previous cycle,
the produced cache data is discarded,
the correct hash key is fed back to update in the prediction table (by setting the $update$ flag),
and the corresponding cache access is replayed (by setting the $replay$ flag),
which incurs a penalty of two cycles in minimum.

\subsection{A speculative L1-I cache}

In the L1-D cache, the target VA comes from a register inside the BOOM load store unit~\cite{Celio2015}.
Adding the prediction table does not make the path critical in timing.
However, the target VA in the L1-I cache might come from
the fully-associative branch target buffer (BTB)~\cite{Celio2015}.
The path is already tight in timing.
To avoid prolonging the clock period,
the implementation of \textsf{RCL} in the L1-I cache can introduce further speculation.

\begin{figure}[bt]
\centering{
\includegraphics[width=0.43\textwidth]{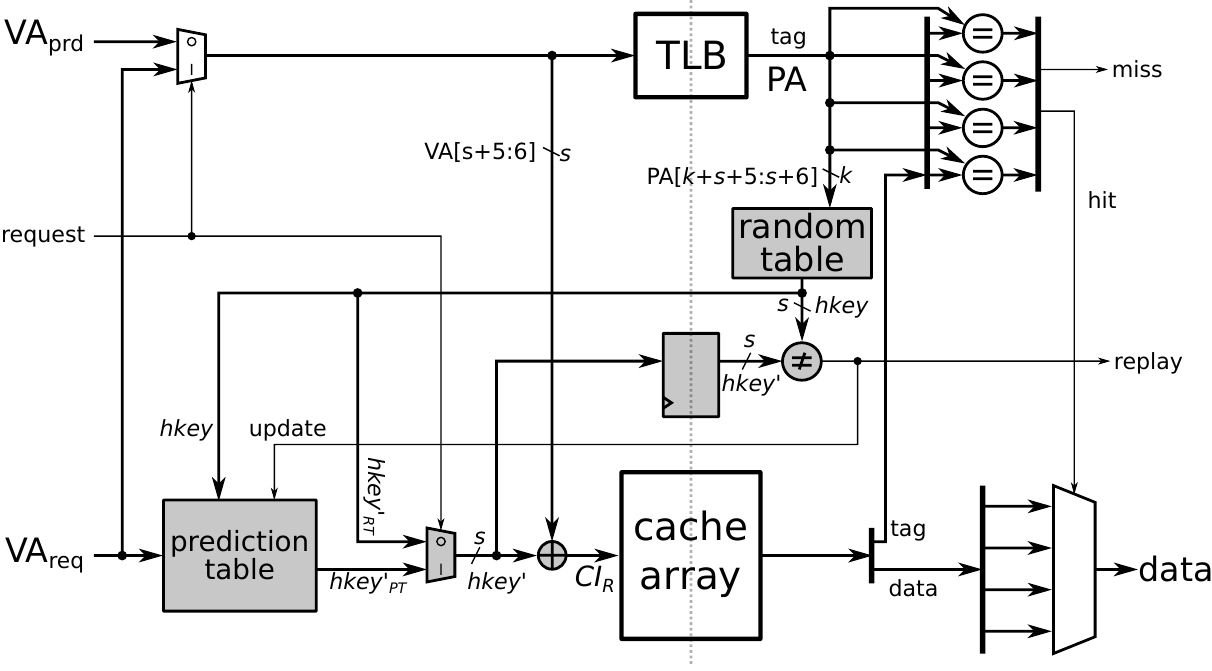}
\vspace{-0.8em}
}
\caption{A speculative \textsf{RCL} L1-I cache}\label{fig_rcl_l1i_speculate}
\vspace{-0.5em}
\end{figure}

Figure~\ref{fig_rcl_l1i_speculate} reveals
the internal structure of the speculative L1-I cache.
Depending on the sources of requests,
the target VA is separated into the predicted VA ($VA_{prd}$, including the VA generated by the BTB)
and the requested VA ($VA_{req}$, when the predicted VA is wrong or the processor pipeline is redirected).
The timing for the paths though $VA_{prd}$ is much tighter than those through $VA_{req}$.
When $VA_{req}$ is chosen as the target VA (denoted by the positive $request$ flag),
the prediction table is used to generate the speculated hash key ($hkey'_{PT}$)
in the same manner as in the speculative L1-D cache.

When the predicted VA ($VA_{prd}$) is chosen,
the instruction cache always predicts that the current instruction comes from the same page with the previous instruction;
therefore, the hash key of the previous cycle ($hkey$ generated from the random table by the next pipeline stage)
is directly used as the speculated hash key ($hkey'_{RT}$).
The latency incurred by the prediction table is thus avoided.
If this simplified speculation is wrong and $VA_{prd}$ indeed crosses the page boundary,
the hash key check in the next cycle would fail as $hkey'_{RT} \neq hkey$.
$VA_{prd}$ is then thrown into the replay queue and the instruction fetch is replayed using the requested VA $VA_{req}$.
Since the failure of hash key check triggers the update in the prediction table,
the replay is guaranteed with a correct speculation using $hkey'_{PT}$.

We believe the simplified speculation on $VA_{prd}$ is correct most of the time.
The access pattern for instructions is mostly linear.
The instruction flow thus crosses the page boundary for around a thousand instructions.
For branches, BTB is usually accurate only for direct branches while performs poorly for indirect ones~\cite{Chang1997}.
Direct branches are normally used for control flows inside small code segments, such as functions.
The probability of crossing the page boundary is also low.

\subsection{\textsf{RCL} in LLC}

\begin{figure}[bt]
\centering{
\includegraphics[width=0.22\textwidth]{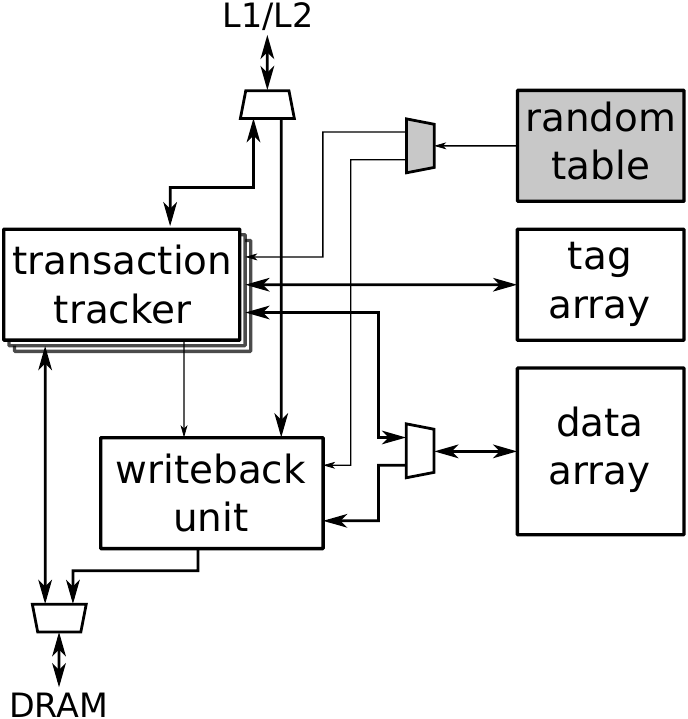}
\vspace{-0.8em}
}
\caption{Implementing \textsf{RCL} in LLC}\label{fig_rcl_llc}
\vspace{-0.5em}
\end{figure}

The cache accesses to LLC are normally treated as simultaneous transactions
and handled by multiple trackers in parallel.
The internal architecture of the L2 (LLC) cache in the BOOM processor is depicted in Figure~\ref{fig_rcl_llc}.
According to Equation~\ref{eqn_ci_rcl_llc},
a random table is added as a shared resource to all trackers and the writeback unit.
Since LLC can tolerate longer access latency than L1 caches,
the random table is implemented in SRAM rather than registers for area efficiency.
The tag and data arrays are indexed by the randomized cache index $CI_R$.
The state machines inside all trackers and the writeback unit are therefore modified according to Equation~\ref{eqn_ci_rcl_llc}.
An arbiter is added to serialize simultaneous accesses to the SRAM-based random table.

\subsection{Initialize random tables}

Random tables are initialized during power-up process or by request.
Individual caches need to guarantee their randomization tables are independently initialized for the best protection.
Here the independence is threefold: the random numbers of the same table are mutually independent,
the random tables in different caches are mutually independently,
and a table is never initialized the same.
If the first independence is violated, indicating the numbers in a table present a pattern,
an attacker might exploit this pattern to decipher the table.
The second independence ensures that the maximum noise is introduced when evicting LLC.
The final independence provides a mitigation against persistent attacks trying to guess the table content.
Similar to the kernel ASLR which is fixed after loading the kernel pages,
it is difficult to dynamically revise the randomization table without causing significant performance degradation.
In the unlikely scenario when part of the randomization table is deciphered by a persistent attack
(probably by a malicious kernel),
a computer system (especially the hypervisor) can reinitialize the randomization table.
Exploring a solution to dynamically update the randomization table with negligible performance impact
is planned as one of our immediate future works.

\begin{figure*}[bt]
\centering{
\subfloat[Execution time overhead compared with the baseline]{\includegraphics[width=0.70\textwidth]{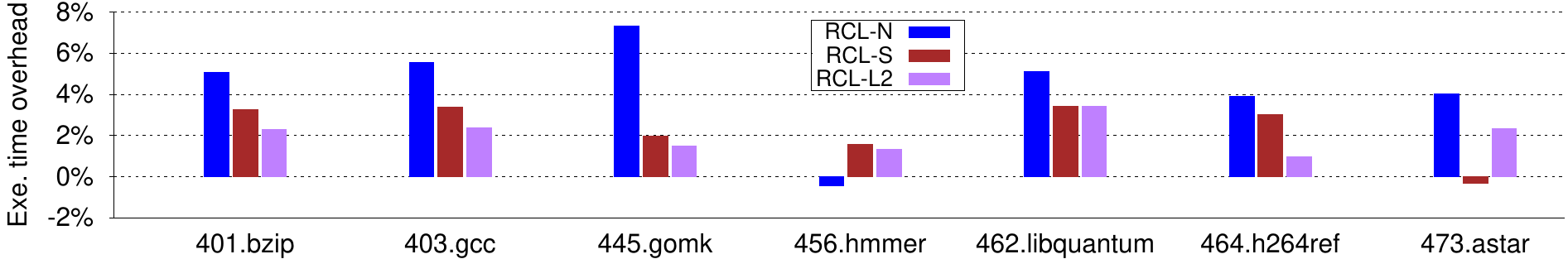}\label{fig_execution_time}} \\
\vspace{-0.5em}
\subfloat[Cache misses per kilo instructions (MPKI)]{\includegraphics[width=0.70\textwidth]{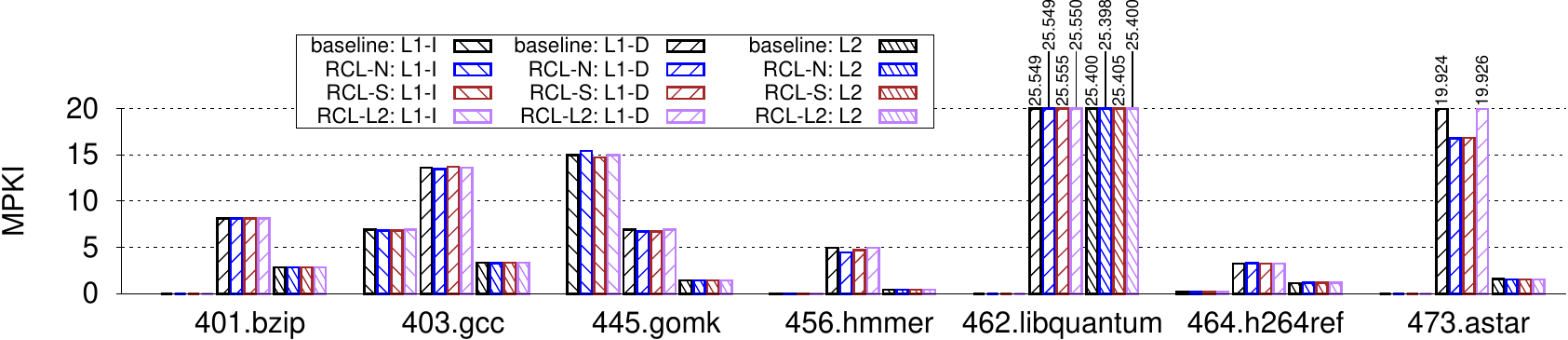}\label{fig_miss_rate}} \\
}
\vspace{-0.8em}
\caption{Running SPEC 2006 on FPGA}\label{fig_fpga_spec}
\vspace{-0.5em}
\end{figure*}

To meet these requirements,
a production-ready cache implementation should
use a true random number generator (TRNG)~\cite{Dichtl2007,Van2012}.
The TRNG can be shared between multiple caches to reduce the area overhead,
as long as tables are initialized independently.
After all tables are initialized, they remain constant during normal cache operations.
The implementation of TRNG does not affect the timing of \textsf{RCL}.
The actual design of TRNG is thus outside the scope of this paper.
In our implementation, simple pseudo-random number generators are used
to initialize the randomization tables.

\section{Evaluation}\label{sec_perf}

\subsection{Experiment Platforms}\label{subsec_platform}

Four different designs of BOOM SoCs have been implemented and evaluated with various configurations:
\begin{itemize}[noitemsep,nolistsep]
\item \emph{Baseline}: The original BOOM SoC.
\item \emph{RCL-N}: All caches are implemented with the non-speculative version of \textsf{RCL}.
\item \emph{RCL-S}: All caches are enabled with \textsf{RCL} but the speculative \textsf{RCL} is implemented in L1 caches.
\item \emph{RCL-L2}: \textsf{RCL} is enabled only in L2 (LLC).
\end{itemize}
Recalling the analysis related to
the automatic creation of eviction sets in Section~\ref{subsec_ana_c}.C (Figure~\ref{fig_eviction_set_rcl}),
applying \textsf{RCL} to LLC alone is effective to thwart most cross-core conflict-based attacks.
Considering flushing L1 caches during process switching~\cite{Domnitser2012, Zhang2013a}
is a strong protection against cross-process attacks targeting L1 caches,
some processors might choose to apply \textsf{RCL} only to LLC as in \emph{RCL-L2}.
The default parameters shared by all BOOM SoCs are illustrated in Table~\ref{tab_boom_param}.

\begin{table}[bt]
  \caption{Parameters of the BOOM SoC}\label{tab_boom_param}
  \begin{center}  
  \small{
  \begin{tabular}{lc}
    \toprule
      Description               & Default value \\
    \midrule
      ISA                       & 64-bit RISC-V \\
      Pipeline stages           & 6             \\
      Issue width               & 4             \\
      Commit width              & 2             \\
      TLB entries               & 8             \\
      Reorder buffer entries    & 48            \\
      L1 cache (I \& D)         & $32KB$ (8-way)  \\
      L2 cache                  & $1MB$ (16-way)  \\
      Clock frequency (FPGA)    & $50MHz$       \\
      Clock frequency (ASIC)    & $750MHz$      \\
    \bottomrule
  \end{tabular}
  }
  \end{center}
  \vspace{-0.5em}
\end{table}

\begin{figure*}[bt]
\centering{
\vspace{-0.5em}
\subfloat[Execution time overhead compared with the baseline]{\includegraphics[width=0.70\textwidth]{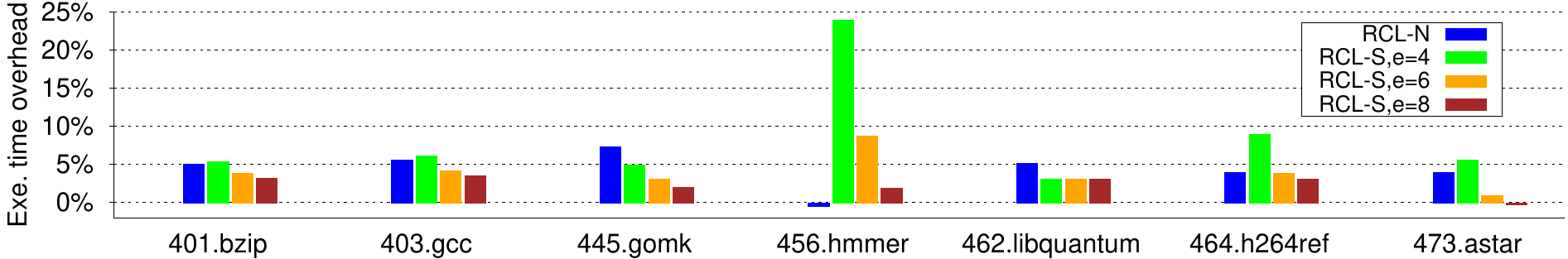}\label{fig_speculation_execution_time}} \\
\vspace{-0.5em}
\subfloat[Cache/PT/TLB misses per kilo instructions (MPKI)]{\includegraphics[width=0.70\textwidth]{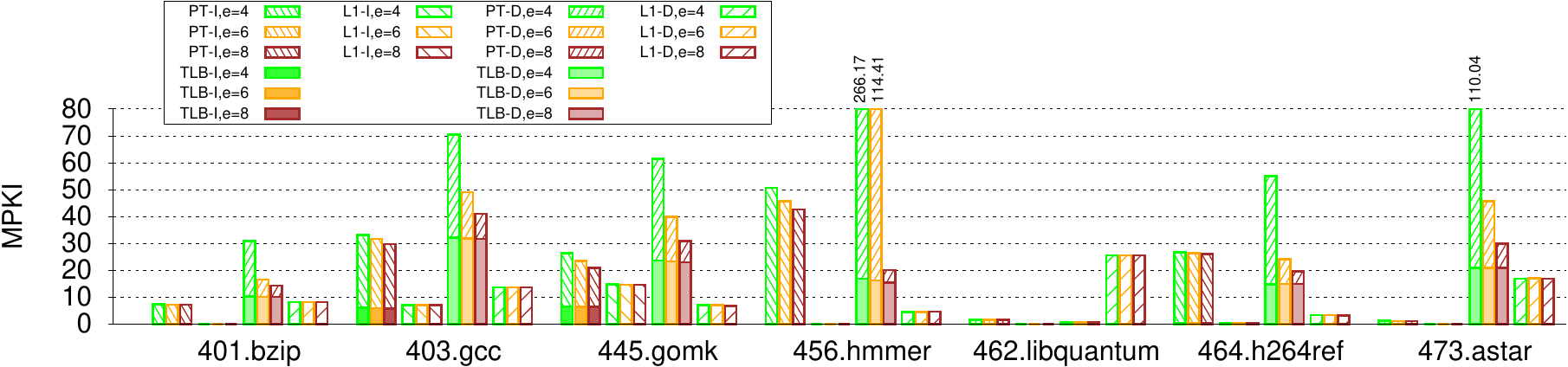}\label{fig_speculation_miss_rate}} \\
}
\vspace{-0.8em}
\caption{Impact of speculation on performance}\label{fig_fpga_speculation}
\vspace{-0.5em}
\end{figure*}

All SoCs are tested on a Xilinx ZC706 FPGA~\cite{Xilinx954}.
We have managed to run a subset of the SPEC 2006 benchmark suite~\cite{Henning2006}.
Since the RISC-V ISA and its Linux support are in constant development,
a number of benchmark cases fail to compile or crash in execution.
These failed cases are not caused by \textsf{RCL} because
all benchmark cases that successfully run on the original BOOM SoC also run on the ones supporting \textsf{RCL}.

To evaluate the area overhead of supporting \textsf{RCL} in ASICs,
all processor designs has been implemented using a standard cell library
and a memory compiler in a 40~nm technology.
The Synopsys Design Compiler topological synthesis flow is used
to provide accurate post-synthesis area and frequency figures.
All processors are constrained at 750~MHz for area results
while they are pushed to their limits for frequency results.

\subsection{Running the BOOM SoCs on FPGAs}\label{subsec_fpga}

Figure~\ref{fig_execution_time} reveals the prolonged execution time in supporting \textsf{RCL} in BOOM SOCs.
The average execution time increases by 4.4\%, 2.3\% and 2.0\% for RCL-N, RCL-S and RCL-L2 respectively.
The execution time increases for two reasons:
The L1 cache access latency increases due to the extra cycle in RCL-N or misprediction in RCL-S.
The L2 cache access latency increases because extra cycles are needed to access the randomization table.
In all benchmark cases, RCL-N incurs the longest execution time as it suffers both latency overheads.
Utilizing speculation significantly reduces the latency overhead in L1 caches.
The speculative RCL-S present the similar execution time overhead with the L2 only implementation (RCL-L2).
In all benchmark cases, the execution time is prolonged for less than 3.5\% for both RCL-S and RCL-L2.

Note that not all cases suffer from prolonged execution time.
For certain combinations, such as in `456.hmmer' (with RCL-N) and `473.astar' (with RCL-S),
the regular access pattern of an application might cause extra conflict misses.
\textsf{RCL} might provide a balanced data distribution which reduces conflict misses,
resulting in a slightly reduced execution time.

For most applications, \textsf{RCL} should incur only marginal impact on the cache miss rate.
To verify this assumption by real hardware,
hardware performance counters are added to all levels of caches.
These counters constantly record the numbers of cache accesses, cache misses and write-backs
by monitoring the cache control logic.
Figure~\ref{fig_miss_rate} reveals the miss per kilo instructions (MPKI) collected from all cache levels.
According to the results,
there is no significant change in the MPKI on both cache levels with or without \textsf{RCL}
except for `473.astar' where supporting \textsf{RCL} actually reduces the MPKI in L1-D by 16\%.
To be specific, the average MPKI in L1-I is increased by 1.6\% for RCL-N and decreased by 1.7\% for RCL-S.
The average MPKI in L1-D is reduced by 4.9\% and 4.3\% for RCL-N and RCL-S respectively.
The change of the average MPKI in all other combinations are less than 0.1\%.

To analyze the efficiency of the speculation,
Figure~\ref{fig_fpga_speculation} reveals the impact of
the number of entries inside the prediction table (PT)
on the execution time and cache misses.
The number of PT entries $e$ increases from 4 to 8,
which is the maximum number of entries without affecting the frequency.
As shown in Figure~\ref{fig_speculation_execution_time},
most benchmark cases are sensitive to the number of PT entries except for `462.libquantum'.
`456.hmmer' is the most sensitive case
where reducing $e$ to 4 (RCL-S,e=4) leads to >20\% execution time overhead compared with (RCL-S,e=8).
Nevertheless, the execution time overhead for all benchmark cases is reduced to 4.2\% if $e \ge 6$.
On average, the execution time increases by 8.4\%, 4.0\% and 2.3\%
for 4, 6 and 8 PT entries respectively.

Figure~\ref{fig_fpga_speculation} shows the MPKIs in PTs, TLBs and the L1 caches.
It is obvious that the size of PT has no impact on the MPKIs of TLBs and caches.
The extra execution overhead is caused by the misprediction inside PT.
This effect becomes more evident when a benchmark case lacks for spatial locality in its data accesses,
such as `403.gcc', `456.hmmer' and `473.astar'.
In these cases, the data TLB has a higher MPKI than the data cache itself.
For the case of `456.hmmer', which shows the most sensitivity to speculation,
although the instruction shows a strong spatial locality,
the pattern of heavily code-reuse leads to frequency cross-page branches,
which leads to significant amount of misprediction in the simplified speculation
for the $VA_{prd}$ produced by BTB (as shown in Figure~\ref{fig_rcl_l1i_speculate}). 

\subsection{Hardware Overhead in ASICs}

To provide a relatively accurate estimation on supporting \textsf{RCL} in ASIC implementations,
all configurations of the BOOM SoCs have been synthesized using the SMIC 40~nm technology.
The area results of all levels of caches with different cache configurations are listed in Table~\ref{tab_asic_area}
where the area results of the caches supporting \textsf{RCL} are normalized.
In all caches, the parameter $k$ is set to $s$ by default, leading to a random table of $2^s$ entries.
As described in Section~\ref{subsec_ana_a}.A,
a cache can increase the random table ($w \cdot 2^{k+s+6} > 2^{21}$)
to prevent an attacker from deterministically create eviction sets using a larger page.
For each L1 configuration, the area overhead with an enlarged random table is also evaluated (with a larger $k$).
The L2 cache is normally large enough to cover a larger page.

\begin{table}[bt]
  \caption{Cache Area}\label{tab_asic_area}
  \begin{center}  
  \small{
  \begin{tabular}{lccccc}
    \toprule
              & Size   & Config          & Baseline &    RCL-N    & RCL-S \\
              & KiB    & $(w,s,k)$       & mm$^2$   &     \%      &  \%   \\
    \toprule
    \multirow{6}{*}{L1-I}
              & \multirow{4}{*}{16}
                       & $(4,6,6)$      &  \multirow{2}{*}{0.260} &    1.7   &  2.0 \\
              &        & $(4,6,8)$      &                         &    4.6   &  4.9 \\
    \cmidrule{3-6}
              &        & $(8,5,5)$      &  \multirow{2}{*}{0.377} &    0.7   &  0.9 \\
              &        & $(8,5,9)$      &                         &    4.5   &  4.7 \\
    \cmidrule{2-6}
              & \multirow{2}{*}{32}
                       & $(8,6,6)$      &  \multirow{2}{*}{0.458} &    0.9   &  1.1 \\
              &        & $(8,6,8)$      &                         &    2.6   &  2.8 \\
    \toprule
    \multirow{6}{*}{L1-D}
              & \multirow{4}{*}{16}
                       & $(4,6,6)$      &  \multirow{2}{*}{0.400} &    1.6   &  2.0 \\
              &        & $(4,6,8)$      &                         &    3.4   &  3.8 \\
    \cmidrule{3-6}
              &        & $(8,5,5)$      &  \multirow{2}{*}{0.518} &    0.9   &  1.3 \\
              &        & $(8,5,9)$      &                         &    3.8   &  4.2 \\
    \cmidrule{2-6}
              & \multirow{2}{*}{32}
                       & $(8,6,6)$      &  \multirow{2}{*}{0.602} &    1.1   &  1.4 \\
              &        & $(8,6,8)$      &                         &    2.3   &  2.6 \\
    \toprule
    \multirow{8}{*}{L2}
              & \multirow{2}{*}{256}
                       & $(8,9,9)$     &  0.948                  &    \multicolumn{2}{c}{3.4} \\
              &        & $(16,8,8)$    &  1.019                  &    \multicolumn{2}{c}{2.6} \\
    \cmidrule{2-6}
              & \multirow{2}{*}{512}
                       & $(8,10,10)$  &  1.801                  &    \multicolumn{2}{c}{2.2} \\
              &        & $(16,9,9)$    &  1.860                  &    \multicolumn{2}{c}{1.7} \\
    \cmidrule{2-6}
              & \multirow{2}{*}{1024}
                       & $(8,11,11)$   &  3.550                  &    \multicolumn{2}{c}{1.3} \\
              &        & $(16,10,10)$  &  3.566                  &    \multicolumn{2}{c}{1.1} \\
    \cmidrule{2-6}
              & \multirow{2}{*}{2048}
                       & $(8,12,12)$   &  7.032                  &    \multicolumn{2}{c}{1.0} \\
              &        & $(16,11,12)$  &  7.063                  &    \multicolumn{2}{c}{0.6} \\
    \bottomrule
  \end{tabular}
  }
  \end{center}
\vspace{-0.5em}
\end{table}

The register-implemented random table with the default $k$ ($k=s$) incurs a marginal area overhead in all L1 caches.
The smallest cache (4-way 64-set) suffers the largest area increase of only 2.0\%.
Even with an enlarged random table, the area overhead is less than 5\% for all caches.
The speculation circuit in RCL-S introduces a negligible area overhead of around 0.2$\sim$0.4\% for all caches.
In L2 caches, the SRAM-implemented table shows a good scalability with the increasing number of sets.
The smallest L2 cache (8-way 512-set) has the maximal area overhead of only 3.4\%.

\begin{figure}[bt]
\centering{
\subfloat[L1 freq. vs. RT sizes ($2^k$)]{\includegraphics[width=0.25\textwidth]{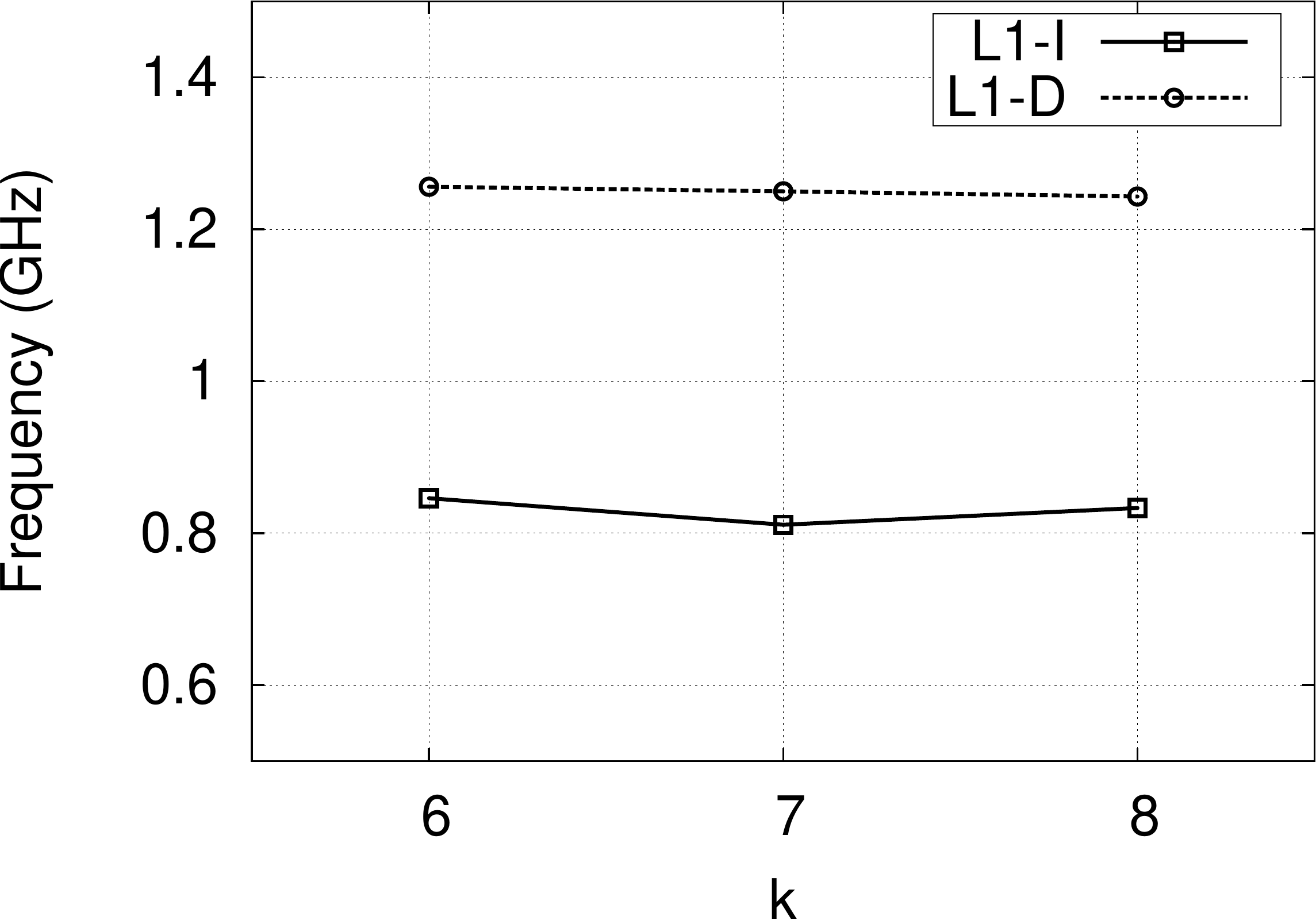}\label{fig_l1_freq}} \\
\vspace{-0.5em}
\subfloat[L2 freq. vs. cache sizes]{\includegraphics[width=0.25\textwidth]{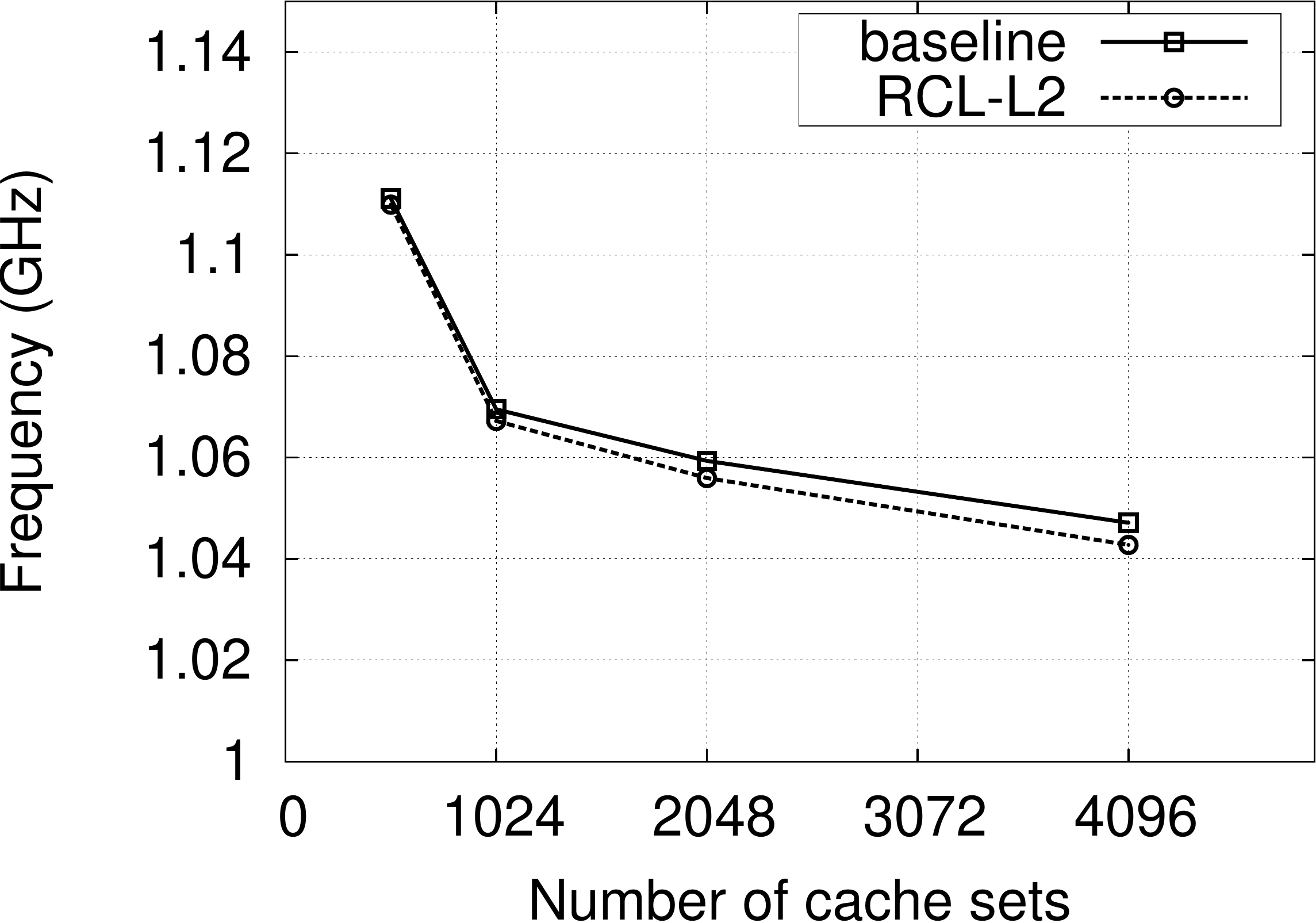}\label{fig_l2_freq}}
}
\vspace{-0.8em}
\caption{Cache frequency}\label{fig_freq}
\vspace{-0.5em}
\end{figure}

As for the impact on clock frequency, L1 caches runs at the same frequency with the processor core.
L1-I has a tighter timing than L1-D.
The baseline BOOM SoC can run at 843~MHz.
RCL-N reduces the frequency to 756~MHz introducing a drop of 10.3\%.
To reduce this effect, the speculative RCL-S runs at 846~MHz, the same speed with the baseline.
Figure~\ref{fig_l1_freq} reveal
the clock frequencies of L1 caches with different random tables in a RCL-S BOOM SoC.
Thanks to the speculation circuit, the random table is moved of the critical path.
The size of the random table has no significant impact on clock frequency.
It is also clear that the L1-I has a much tighter timing compared with L1-D.
For RCL-L2, which implements \textsf{RCL} only in the L2 cache,
Figure~\ref{fig_l2_freq} reveals the frequency drop with the enlarged cache and random table.
Supporting \textsf{RCL} in L2 incur a negligible additional drop in frequency from 0.11\% for the 512-set cache to 0.42\% for the 4096-set cache.

\section{Related Work}\label{sec_related}

As described in Section~\ref{sec_defense}, \textsf{RCL} is not the first defense trying to randomize the cache layout.
RPcache~\cite{Wang2007} and the later NewCache~\cite{Wang2008}
are the first to randomize the mapping from VAs to cache sets.
The randomized cache index in the PRcache can be described as
$f(VA[s+5:6])$ where $f()$ is a random function similar to the random table in \textsf{RCL}.
This defense can effectively thwart conflict-based attacks infer information from the L1 cache indexes.
However, it does not alter the fact
that congruent cache lines (sharing $VA[s+5:6]$) are still mapped to the same cache set.
The cache layout in L2/LLC is not randomized.
These two drawbacks make PRcache ineffective in front of the advanced side-channel attacks typed \textbf{C} and \textbf{D}.
Although later the NewCache enhanced the randomization by a ReMapping table~\cite{Wang2008}, the aforementioned issues remains.
Both PRcache and NewCache require software to set security flags on cache lines,
which unnecessarily introduces modifications in software.
Introducing PA into the cache layout randomization,
\textsf{RCL} is the first pure hardware defense
that deliberately avoids storing congruent cache lines in the same cache set,
applicable in all cache levels and require no software involvement.

Any cache related technologies that introduce randomness into the cache layout
can potentially work with \textsf{RCL} and strengthen its defense.
Introducing randomness into the hardware prefetcher~\cite{Fuchs2015}
or allow cache lines to decay at random intervals~\cite{Keramidas2008}
provide extra defense against the flush-based attacks.
If the secrity-critical data can be identified by software,
such as the lookup tables normally used in cryptographic algorithms,
randomizing the fetched cache line~\cite{Liu2014}
or ask software to explicitly randomize the security-critical data before storing them to memory~\cite{Schwarz2018}
are effective as well.
The complex addressing scheme utilized in modern Intel processors~\cite{Maurice2015}
can be enhanced to provide run-time slice mapping.
The skewed cache~\cite{Seznec1993}, which was proposed to improve cache associativity,
and the compressed cache~\cite{Sardashti2014} can be applied along with \textsf{RCL}
to introduce further randomness into the cache layout.

Combining \textsf{RCL} with other defenses provides further protection.
All the defenses described in Table~\ref{tab_rcl_protection} naturally work along with \textsf{RCL},
such as constant time programming~\cite{Bernstein2012},
flushing of local caches during process switching~\cite{Domnitser2012, Zhang2013a},
KPTI ~\cite{Gruss2017} and TEE~\cite{McKeen2013, Costan2016}.
For cache partitioning, \textsf{RCL} can work with all defenses using way partitions~\cite{Domnitser2012, Yan2017}.
However, cache partitioning based on page color~\cite{Shi2011} and explicit cache management~\cite{Kim2012, Liu2016}
may no longer work
because \textsf{RCL} has randomized the cache layout
and cache lines with different page colors can be stored in the same cache set.

\section{Conclusion}

Based on a systematic review of the common cache-based side-channel attacks and
the existing defenses, automatic side-channel attacks and bypassing the user ASLR
have been identified as largely undefended in current computer systems.
A pure hardware defense against a wide range of conflict-based side-channel attacks,
namely Remapped Cache Layout (\textsf{RCL}), is proposed.
\textsf{RCL} deliberately randomizes the mapping from addresses to cache sets.
It prevents attackers from accurately infer the location of her data in caches
or using a cache set to infer her victim's data.
To our best knowledge, \textsf{RCL} is the first defense to thwart
the automatic side-channel attacks and the attacks of bypassing the user ASLR.
It is a also pure hardware defense which requires no software involvement.
\textsf{RCL} has been implemented in all levels of caches into the BOOM SoC.
A speculative cache structure is proposed to reduce the cache access latency in \textsf{RCL} enabled L1 caches.
Collected from running the SPEC 2006 benchmark on the \textsf{RCL} enabled BOOM in an FPGA,
the detailed evaluation results show that \textsf{RCL} incurs only small costs in area, frequency and execution time.

%-------------------------------------------------------------------------------
\section*{Acknowledgments}
%-------------------------------------------------------------------------------
This work was supported by the CAS Pioneer Hundred Talents Program,
the National Natural Science Foundation under grant No. 61802402,
and internal grants from the Institute of Information Engineering, CAS.

%%%%%%%%% -- BIB STYLE AND FILE -- %%%%%%%%
\bibliographystyle{ieeetr}
\bibliography{ref}
%%%%%%%%%%%%%%%%%%%%%%%%%%%%%%%%%%%%

\end{document}